\documentstyle[preprint,eqsecnum,aps]{revtex}
\begin{document}
\draft 
\title{\Large The Perturbative Gross Neveu Model 
Coupled to a Chern-Simons Field: A Renormalization Group Study}
\author{V. S. Alves\cite{byline}, M. Gomes, S. V. L. Pinheiro\cite{byline}
and  A. J. da Silva}
  \address{Instituto de F\'\i sica, USP\\
 C.P. 66318 - 05315-970, S\~ao Paulo - SP, Brazil}
\date{1998}

\maketitle

\begin{abstract}
  In 2+1 dimensions, for low momenta, using dimensional
  renormalization we study the effect of a Chern-Simons field on the
  perturbative expansion of fermions self interacting through a Gross
  Neveu coupling. For the case of just one fermion field, we verify
  that the dimension of operators of canonical dimension lower than
  three decreases as a function of the Chern-Simons coupling.

\end{abstract}
\section{Introduction}
Effective field theories is a subject of great interest in theoretical
physics not only due to their potential applications but also because
they provide new insights into the way we look at field theories
\cite{Wi}. From this perspective nonrenormalizable models have
acquired a new status as they may become physically relevants at low
energies \cite{Ba}. The point is that, if the scale of energy one is
interested is low enough, the ambiguities due to the virtual states of
high energy do not show up or, equivalently, are not meaningful. On
the energy interval where this happens the theory proceeds as an usual
renormalizable one.  Nonetheless, as observed in \cite{Manohar}, the
use of a mass independent regularization is almost mandatory to
guarantee that high order counterterms can be effectively neglected.

The ultraviolet behavior of the Green functions may be changed by a
rearrangement of the perturbative series. In fact, the incorporation
of vacuum polarization effects in general improves the convergence
properties of the resumed series; this mechanism is well known to be
operative in the context of the $1/N$ expansion. In particular, Gross
Neveu \cite{Gross} or Thirring \cite{Gomes} like four fermion
interactions which in (2+1) dimensions are perturbatively
nonrenormalizable become renormalizable within the framework of the
$1/N$ expansion \cite{Gross}. This result has motivated a series of
investigations on the properties of these theories \cite{Hands}.  In
particular, using renormalization group (RG) methods, it has been
proved that the $N$ component Gross Neveu model in 2+1 dimensions is
infrared stable at low energies but has also a nontrivial ultraviolet
stable fixed point.  These facts indicate that the theory could be
perturbatively investigated if the momentum is low enough. This
actually would be the only remaining possibility for small $N$.
  
It has recently been conjectured that in 2 +1 dimensions, besides the
$1/N$ expansion, there is another way to improve the ultraviolet
behavior of Feynman amplitudes.  By coupling fermion fields to a
Chern-Simons field the scale dimension of field operators could be
lowered possibly turning non renormalizable interactions into
renormalizable or, better, super-renormalizable ones.   Using a sharp
cutoff to regulate divergences, this idea was tested in \cite{Chen}
where the effect of the CS field over massless self-coupled fermions
with a quartic, Gross--Neveu like, interaction was studied.
  
In this communication we will pursue this study further by considering
massive fermions and adopting dimensional renormalization
\cite{Giambiagi} as a tool to render finite the Feynman amplitudes. In
this way we evade the ambiguity problem associated with the routing of
the momentum flowing through the associated Feynman graphs
\cite{Willey}. Nevertheless, it should be stressed that our
calculations are valid insofar, as said above, the effect of the
higher order counterterms can be neglected.  Otherwise, new couplings
should be introduced. Our investigations, restricted to the case of
fermions of just one flavor, i. e. $N=1$, show that, differently to
what happens for large $N$,the renormalization group beta function has
only a trivial infrared stable fixed point.  Moreover, the operator
dimensions of the basic field and composites of canonical dimension
lower than three are monotonic decreasing functions of the CS
parameter.  This indicates that the Feynman amplitudes have a better
ultraviolet behavior if the underlying theory is renormalizable.
However, no improvement in the ultraviolet behavior seems to occur if
the composite operators have canonical dimension bigger than three.
 
Our work is organized as follows. In section II some basic properties
of the model as Feynman rules, ultraviolet behavior of Feynman
diagrams and comments on the regularization procedure are presented.
The derivation and calculation of the renormalization group parameters
are indicated in section III. Section IV contains a discussion of our
results as well as our conclusions.  Details of the calculations of
the pole part of the relevant amplitudes are described in the
appendices \ref{appendixA} and \ref{appendixB}.
\section{A quartic interaction}
We consider a self-interacting two--component spinor field minimally
coupled to a CS field. The Lagrangian density
is given by
\begin{equation}
\label{1}
{\cal L} = \frac{1}{2\pi\alpha}
\epsilon^{\mu\nu\alpha}\,\partial_{\mu} A_\nu \,A_{\alpha}
+\bar{\psi}(i \not \! \partial-m) \psi+ \bar \psi\gamma^\mu \psi A_\mu
- G(\bar \psi\psi)(\bar \psi\psi) + \frac{1}{2\lambda}(\partial_\mu A^\mu)^2,
\end{equation}

\noindent
The Dirac field $\psi_{\alpha}$ represents particles and
anti-particles of spin up and the same mass $m$ (the parameter m is to
be taken positive) \cite{o1}.  The Gross-Neveu term in (\ref{1}) is
the most general Lorentz covariant quartic self-interaction, for the
Thirring-like vector interaction is not independent but satisfies
$:(\bar \psi\gamma^\mu \psi)(\bar \psi \gamma_\mu\psi):= -3:(\bar \psi
\psi)(\bar \psi \psi): $.  $\lambda$ is a gauge fixing parameter but
for simplicity we will always work in the Landau gauge, formally
obtained by letting $\lambda\rightarrow 0$.  In this gauge, the Green
functions may be computed using the Feynman rules depicted in
Fig. \ref{fig1}. For convenience, we have introduced auxiliary dotted
lines, hereafter called auxiliary GN lines, to clarify the structure
of the four fermion vertex.

Divergences show up, the degree of superficial
divergence of a generic graph $\gamma$ being
\begin{equation}
d(\gamma)= 3 - N_A - N_F +V,
\end{equation}
where $N_A$ and $N_F$ are the number of external lines associated with
the propagators for the the Chern--Simons and the fermion fields,
respectively; $V$ denotes the number of quartic vertices in $\gamma$.
The model is of course nonrenormalizable and the number of
counterterms necessary to render the amplitudes finite increases with
the order of perturbation but, to a given order, the number of
counterterms is finite. To do calculations we will employ dimensional
renormalization starting at the space-time dimensio $d$. It is
therefore convenient to introduce a dimensionless coupling $g$ and a
renormalization parameter $\mu$ through $G= (g/\Lambda)
\mu^{\epsilon}$ and $\alpha \rightarrow \alpha \mu^\epsilon$, where
$\epsilon= 3 -d$ must be set zero at the end. The massive parameter
$\Lambda$ must be considered much bigger than any typical momenta and
than the fermion mass $m$; it sets the scale which limits the region
where our results are valid.  Divergences will appear as poles in
$\epsilon$ and a renormalized amplitude is given by the $\epsilon$
independent term in the Laurent expansion of the corresponding
regularized integral.  However, at one loop level no infinities will
remain after the remotion of the regulator. This is so because the
poles for a graph $\gamma$ may occur only at even values of the degree
of superficial divergence of $\gamma$\cite{Speer}.  Moreover, it is
easy to check that asymptotically, i. e., for zero external momenta,
one loop graphs with even degree of divergence are odd functions of
the loop momentum and therefore vanishes, after symmetric integration.

\section{Renormalization Group}

As known, Green functions of renormalizable models, which have been
made finite by the subtraction of pole terms in the dimensionally
regularized amplitudes, satisfy a 't Hooft-Weinberg type
renormalization group equation \cite{Thooft}.  Nonrenormalizable
models require special consideration since the form of the effective
Lagrangian changes with the order of perturbation. However, at
sufficient small momenta, such that the effect of the new counterterms
may be neglected, the Green functions will still approximately satisfy
the RG equation.  Thus, although being nonrenormalizable, for small
enough momenta, the Green function of the theory (\ref{1}) satisfy the
following renormalization group equation
\begin{equation}\label{2}
[\Lambda \frac{\partial\phantom {a}}{\partial \Lambda} + \mu 
\frac{\partial\phantom {a}}{\partial \mu}+ \delta \, m 
\frac{\partial\phantom {a}}{\partial m}+
\beta \frac{\partial\phantom{a}}{\partial g}- N \gamma]\Gamma^{(N)}(p_1, 
\ldots p_N)\approx 0 ,
\end{equation}
where $\Gamma^{(N)}(p_1,\ldots p_N)$ denotes the vertex function of
$N$ fermion fields (since $A_\mu$ is not a dynamical field we shall
not consider vertex functions having external vector fields). The
symbol $\approx$ means equality in the region where all counterterms
different from those terms already present in (\ref{1}) can be
neglected.  As a consequence of the Coleman--Hill theorem, which
states that all radiatives corrections to the CS term are finite,
\cite{Coleman}, the $beta$ function for the CS coupling $\alpha$
vanishes identically; that explains why the term with a derivative
with respect to $\alpha$ is absent from (\ref{2}).

The coefficients $\delta$, $\beta$ and $\gamma$ in equation (\ref{2})
may be obtained by formally computing the action of the differential
operator over the two point and four point Green functions. For the
two point function, up to second order in the coupling constants, we
have
\begin{eqnarray}\label{3}
\Gamma^{(2)}(p)&=& i (\not \! p - m) + \frac{g}{\Lambda} 
\mu^\epsilon I^{(2)}_1+
 \alpha \mu^\epsilon I^{(2)}_2 +\alpha^2 (1-{\cal T})\mu^{2\epsilon} 
I^{(2)}_3  \nonumber \\
&\phantom a &+\frac{g  \alpha}{\Lambda} (1-{\cal T})
\mu^{2 \epsilon} I^{(2)}_4 +
\frac{g^2}{\Lambda^2}(1-{\cal T})\mu^{2\epsilon} I^{(2)}_5,
\end{eqnarray}

\noindent
where the limit $\epsilon \rightarrow 0$ must be understood. In the
above expression $I^{}_i$, $i=1,\ldots,5$, denote the regularized
Feynman amplitudes. In particular, the graphs ascribed to $I_3$, $I_4$
and $I_5$ have been depicted in the figures \ref{fig2}, \ref{fig3} and
\ref{fig4}, respectively; ${\cal T}$ is an operator to remove the pole
term in the amplitudes to which it is applied. As mentioned earlier,
the amplitudes $I^{(2)}_{1}$ and $I^{(2)}_2$ which are associated with
one loop diagrams are finite.  Inserting (\ref{3}) into (\ref{2})
allows us to determine the coefficients $\delta $ and $\gamma$ as
follows.  Initially notice that as $\Lambda$ enters into the
perturbative expansion only in the combination $g/\Lambda$, fixing
$\beta$ in lowest order as being equal to $g$ eliminates all
contributions of the term with the derivative with respect to
$\Lambda$ in (\ref{2}). After that, up to the order we will study, in
(\ref{2}) there will be no mixing of higher order contribution to
$\beta$ with those to $\gamma$ and $\delta$.

Using the expansions,
\begin{eqnarray}\label{4a}
\delta &=& \sum_{i,j} \delta_{i,j} g^i \alpha^j, \\
\gamma &=& \sum_{i,j} \gamma_{i,j} g^i \alpha^j,\label{4b} 
\end{eqnarray}
where the sum is restricted to $i+j\leq 2$, we get
\begin{eqnarray}\label{5}
\delta_{1,0}&=& \delta_{0,1}= \gamma_{1,0}= \gamma_{0,1}=0,\\
\delta_{0,2} &=& - 2 i (A_3+ B_3)\qquad \phantom {ab}\gamma_{0,2}=-i B_3\\
\delta_{1,1} &=&- \frac{2m}{\Lambda} (A_4+ B_4)\qquad \phantom {abc}
\gamma_{1,1}=-\frac{m}{\Lambda} B_4 \\
\delta_{2,0} &=& \frac{2 im^2}{\Lambda^2} (A_5+ B_5)\qquad  \gamma_{2,0}=
\frac{i m^2}{\Lambda^2} B_5\\
\end{eqnarray}

\noindent
where $A_i$ and $B_i$, for $i=3,4,5$ are defined by writing the pole term
for the amplitude $I^{(2)}_i$  as 
\begin{eqnarray}\label{6}
\mbox{Pole term of}\quad I^{(2)}_3 &=&( m A_3 + \not 
\! p B_3)\frac{1}{\epsilon}\\
\mbox{Pole term of}\quad I^{(2)}_4 &=& -i(m^2 A_4+ m \not 
\! p B_4 + O(p^2))
\frac{1}{\epsilon}\\
 \mbox{Pole term of}\quad I^{(2)}_5 &=& -(m^3 A_5+ m^2 \not 
\! p B_5 + O(p^2)) 
\frac{1}{\epsilon}
\end{eqnarray}

The appendix \ref{appendixA} presents a detailed analysis of the
various contributions to these parameters. From (\ref{I15}),
(\ref{I18}) and (\ref{I20}), the final result is
\begin{eqnarray}
\delta & =& - \frac{8}{3}\alpha^2-\frac{11}{8\pi}\frac{m}{\Lambda}g \alpha+
\frac{7}{12\pi^2}\frac{m^2}{\Lambda^2} g^2 \\
\gamma &=& -\frac{1}{12} \alpha^2 - \frac{1}{8\pi}\frac{m}{\Lambda} g \alpha+
\frac{5}{48 \pi^2}\frac{m^2}{\Lambda^2} g^2
\end{eqnarray}

\noindent
Letting $m \rightarrow 0$, we note that our determination of $\gamma$
agrees with \cite{Chen}. In the region $m \ll \Lambda$, however, our
result presents corrections for nonvanishing fermion mass.

To fix $\beta$ we look now at the
four point Green function, which up to third order is (we omit
contributions which by power counting are finite)
\begin{eqnarray}\label{7}
\Gamma^{(4)}(p_1,p_2,p_3,p_4) &=&\mu^{\epsilon}( -i \frac{g}{\Lambda} 
+ \frac{g \alpha}{\Lambda} \mu^{ \epsilon} I^{(4)}_1+ \frac{g^2}{\Lambda^2} 
\mu^{ \epsilon}I^{(4)}_2+ \frac{g}{\Lambda} 
\alpha^2 (1-{\cal T})\mu^{2 \epsilon} I^{(4)}_3 \nonumber \\
&\phantom {a}& + \frac{g^2 \alpha}{\Lambda^2} (1- {\cal T})
\mu^{2 \epsilon} I^{(4)}_4+
\frac{g^3}{\Lambda^3}(1-{\cal T})\mu^{2\epsilon} I^{(4)}_5).
\end{eqnarray}
Here, again, the one loop amplitudes $I^{(4)}_1$ and $I^{(4)}_2$ are
finite because of the use of the dimensional regularization.
Inserting (\ref{7}), the expansion $\beta =\sum_{i,j} \beta_{i,j} g^i
\alpha^j$, and $\delta$ and $\gamma$ given in (\ref{4a}) and
(\ref{4b}) into (\ref{2}), we obtain $\beta_{1,0}=1$,
$\beta_{0,1}=\beta_{1,1}=\beta_{2,0}=\beta_{0,2}=0$ and
\begin{eqnarray}\label{8}
\beta_{1,2}&=&- 4 i B_3- 2  C_3\\
\beta_{2,1} &=&  \frac{2i m}{\Lambda} C_4 - \frac{4 m}{\Lambda} B_4\\
\beta_{3,0} &=& \frac{2 m^2}{ \Lambda^2} C_5 + \frac{4im^2}{\Lambda^2} B_5.
\end{eqnarray}
In the above expressions, $C_i$ for $i=3,4,5$ are related to the pole
part of the amplitudes $I^{(4)}_i$ through, 
\begin{eqnarray}
 \mbox{Pole part of } I^{(4)}_3 &=&
-iC_3/\epsilon ,\\
  \mbox{Pole part of }
I^{(4)}_4 &=& -(mC_4 + O(p))/\epsilon,\\
 \mbox{Pole part of } I^{(4)}_5 &=& i(m^2 C_5 +O(p))/\epsilon .
\end{eqnarray}

\noindent
 In the appendix \ref{appendixB} we have collected the results of the calculations of the pole part of the relevant graphs. Using \ref{B4} we
obtain, finally, 
\begin{equation}\label{8a}
\beta = g + \frac{20}{3} g\alpha^2 - \frac{21m}{2 \pi\Lambda}g^2 \alpha +
\frac{161 m^2}{12 \pi^2\Lambda^2} g^3.
\end{equation}

\section{Discussion and Conclusions}
An inspection of Eq. (\ref{8a}) shows that the renormalization group
{\it beta} function has $g=0$ as a fixed point. As,  for $m\ll \Lambda$, $\beta\approx
g(1+ 20\alpha^2/3 )$ the origin is an infrared stable fixed point.
Actually, this is the only existing fixed point.
Here we are in disagreement with  Ref. \cite{Chen} where a line
of fixed points was found. The diverse conclusions are perhaps due
to the use of different regularizations but a more direct comparison of the
methods
seems unfeasible as the calculations in \cite{Chen} were
not spelled out.

We will examine now the dimensions of some operators. As seen
before  the basic field $\psi$ has operator dimension $d_\psi =1 -\alpha^2/12$,
and so at $g=0$ the Green functions of the fermion field have an improved
ultraviolet behavior as $\alpha$ increases. Similar results are obtained
if one considers composite operators of canonical dimension less than three.
The simplest of them, the mass operator $\bar \psi
\psi$ ha an anomalous dimension given by
\begin{equation}\label{8b}
\gamma_{\bar\psi\psi} = 2 \gamma - 2 Res,
\end{equation}
where $Res$ is the residue coming from graphs contributing to the
vertex function with the insertion of the mass operator, $
\overline \psi\psi$  and having two
external fermionic lines.  For practical purpose this residue may be
computed by taking the mass derivative of the contributions calculated
in the item 1 of appendix \ref{appendixA}. The result is $i A_3 =5/4$.
Thus the dimension of $\bar\psi \psi$ turns out to be equal to
\begin{equation}
d_{\bar \psi\psi}= 2 - \frac{8}{3}\alpha^2.
\end{equation}    

From the computation of the anomalous dimension of the operator
$\bar \psi\psi$ one could easily obtain the dimension of $\bar \psi\not \!\partial \psi$. Indeed, the  dimension of $\bar \psi\not \!\partial \psi$ is given
by  (\ref{8b}) but with the replacement of $Res$ by $3i A_3= 15/4$. Thus
\begin{equation}
d_{\bar \psi\not \partial \psi}= 3 -23/3\alpha^2
\end{equation}
The determination of the anomalous dimension of the operator
$(\bar\psi \psi)^2$ at $g=0$ is more complicated due to the fact that
renormalization in general produces a mixing with other operators of
dimension lower or equal to four. However if we restrict the calculation
to the $m=0$ case, as we will do, only operators of dimension four need
to be considered. A further simplification is obtained by considering
only (formally) integrated operators. We have,
\begin{eqnarray}
\int d^3x N[(\bar \psi \psi)^2] &=& a_1\int d^3 x (\bar\psi \psi)^2  + a_2 
\int d^3 x \bar \psi \partial^2 \psi\\
\int d^3x N[\bar \psi \partial^2 \psi] &=& b_1\int d^3 x (\bar\psi \psi)^2  
+ b_2 \int d^3 x \bar \psi \partial^2 \psi
\end{eqnarray}
where the symbol $N$ indicates a normal product prescription 
corresponding to the subtraction of the pole terms. A direct calculation
gives that $a_2=b_1=0$, $a_1=1+\frac{C_3\alpha^2}{\epsilon}$ and $b_2=1- \frac{\alpha^2}{3\epsilon}$.  A straightforward analysis shows now that the
dimensions of $\int d^3x N[(\bar \psi \psi)^2]$ and $\int d^3x N[\bar \psi \partial^2 \psi]$ are given by
\begin{eqnarray}
&4&+4 \gamma_{02}\alpha^2 - 2 C_3 \alpha^2=4 +\frac{20}{3}
\alpha^2,\\
&4& +2 \gamma_{02}\alpha^2 + \frac{2}{3} \alpha^2= 4+\frac{\alpha^2}{2},
\end{eqnarray} 
respectively. One sees that, at least for the operators that we explicitly
considered, the operator dimension decreases with $\alpha$ accordingly
the canonical dimension is lower or equal to three, in accord with \cite{Chen}. However, if the canonical 
dimension is bigger than three, the operator dimension  increases with
$\alpha$ so that no improvement for nonrenormalizable interactions results. 

Our results are valid if the basic fermion field is flavorless. 
The $N$-flavor case is presently under investigation.

\appendix
\section{}\label{appendixA}
In this appendix we shall present a detailed analysis of the contributions to
the  pole  part of the two point vertex function. Due to the use of the
dimensional regularization the poles only appear at the two loop level,
beginning at second order in the coupling constants. We will examine
separately each order of perturbation, i. e., the orders $\alpha^2$, $g \alpha$ and $g^2$, We have,

1. Order $\alpha^2$.  In this order there are three diagrams
which are shown in Fig. \ref{fig2}. These diagrams give the contributions,

Figure \ref{fig2}$(a):$
\begin{eqnarray}\label{I1}
 I_3(a)&=&4\pi^2 i
\;\epsilon_{\rho\mu\lambda}\;\epsilon_{\nu\theta\alpha}{\cal T}\int
\frac{d^dk_1}{(2\pi)^d}\frac{d^dk_2}{(2\pi)^d}\;k^{\lambda}_{1}\;k^{\alpha}_{1}{\rm Tr}\left[\gamma^{\mu}(\not \!
k_{2}+m)\gamma^{\nu}(\not \! k_{2}-\not \! k_{1}+m)\right]
\nonumber \\
&\times&\frac{\left[\gamma^{\rho}(\not \! p-\not
\!
k_{1}+m)\gamma^{\theta}\right]}{(k^{2}_{2}-m^2)\;[(k_2-k_1)^2-m^2]\;[(p-k_1)^2-m^2]\;(k^{2}_{1})^2}
\nonumber \\
&=&\left(m\;A_3(a)+\not \! p\;B_3(a)\right)\frac{1}{\epsilon},
\end{eqnarray}
where, on the right hand side of the first equality, we have
introduced the operator ${\cal T}$ to extract the pole part of the
expression to which it is applied. From (\ref{I1}) we obtain

\begin{eqnarray}\label{I2}
\frac{A_3(a)}{\epsilon}&=&\frac{1}{2m} {\rm Tr}\;I_3(a)|_{p=0}=\frac{2i\pi^{2}}{m}
{\cal T} \int
\frac{d^dk_1}{(2\pi)^d}\frac{d^dk_2}{(2\pi)^d}
\nonumber \\
&\times&\frac{-8m^3k^{2}_1-8m(k^{2}_1)^2-8mk^{2}_1(k_1.k_2)+8m(k_1.k_2)^2}
{(k^{2}_{2}-m^2)\;[(k_2-k_1)^2-m^2]\;[k^{2}_1-m^2]\;(k^{2}_{1})^2}.
\end{eqnarray}

\noindent
For simplicity, the trace
was taken in the integrand of (\ref{I1}) and calculated directly
at $d=3$. This, of course, does not affect the result for the pole part of
the integrals. Analogously,
\begin{eqnarray}\label{I3}
\frac{B_3(a)}{\epsilon}&=&\frac{1}{2p^2}\;{\rm Tr}(I_3(a)\not \! p)=\frac{2i\pi^{2}}{p^2}{\cal T}\int\frac{d^dk_1}{(2\pi)^d}\frac{d^dk_2}{(2\pi)^d}
\nonumber \\
&\times&\frac{\mbox{Numerator}}{(k^{2}_{2}-m^2)\;[(k_2-k_1)^2-m^2]\;[(p-k^{2}_1)^2-m^2]\;(k^{2}_{1})^2},
\end{eqnarray}
where
\begin{eqnarray}\label{I4}
\mbox {Numerator}&=&-16\epsilon_{\rho\mu\lambda}\;\epsilon_{\nu\theta\alpha}k^{\rho}_1k^{\mu}_2p^{\lambda}k^{\nu}_1k^{\theta}_2p^{\alpha}-16m^2k^{2}_1(k_1\cdot p)-8k^{2}_1(k_1\cdot k_2)(k_1\cdot p)
\nonumber \\
&+& 8m^2(k_1\cdot p)^2+8(k_1\cdot k_2)(k_1\cdot p)^2-8(k_1\cdot p)^2k^{2}_2-8(k_1\cdot k_2)^2p^2+8k^{2}_1k^{2}_2p^2.
\end{eqnarray}

\noindent
Here and in what follows we shall adopt the following procedure for
performing the integrals. We first consider the  $k_2$ integral and use 
Feynman's trick,
\begin{equation}\label{I5}
\frac{1}{a_{1}^{\alpha_1} a_{2}^{\alpha_2}}= \frac{\Gamma[\alpha_1+\alpha_2]}
{\Gamma[\alpha_1]\Gamma[\alpha_2]} \int_{0}^{1} d x \,\frac {x^{\alpha_1-1} (1-x)^{\alpha_2-1}}{[x a_1 + (1-x) a_2]^{ \alpha_1+\alpha_2}}  
\end{equation}
to reduce the denominators containing the variable of integration to only
one denominator. After integrating in $k_2$ we use again Feynman's formula 
(\ref{I5}) to combine the denominators that depend on $k_1$. We then integrate
over $k_1$ and, finally, perform the parametric integrations. In the present
case, after integrating on $k_2$, we get
\begin{eqnarray}\label{I6}
\frac{A_3(a)}{\epsilon}&=&{\cal T}\frac{\pi^{(2-\frac{d}{2})}}{2^{d-3}}\int_{0}^{1}dx\int\frac{d^{d}k_{1}}{(2\pi)^{d}}\;\frac{1}{(k^{2}_{1}-m^{2})\; k_{1}^{2}}
\nonumber \\
&\times&
\left[\Gamma(1-\frac{d}{2})\;\Delta^{\frac{d}{2}-1}+\Gamma(2-\frac{d}{2})\;k^{2}_{1}\;\Delta^{\frac{d}{2}-2}\;(1+x-x^{2})\right]
\end{eqnarray}

and

\begin{eqnarray}\label{I7}
\frac{B_3(a)}{\epsilon}&=&-{\cal T}\frac{1}{p^2}\;\frac{\pi^{(2-\frac{d}{2})}}{2^{d-1}}\int_{0}^{1}dx\int\frac{d^{d}k_{1}}{(2\pi)^{d}}\;\frac{k_1\cdot p}{[(p-k_{1})^2-m^{2}]\;(k^{2}_{1})^{2}}
\nonumber \\
&\times&
\left[4\Gamma(1-\frac{d}{2})\;\left((d-2)(k_1\cdot p)-k^{2}_1\right)\;\Delta^{\frac{d}{2}-1}+8x(1-x)
\right. \nonumber \\
&\times& \left. \Gamma(2-\frac{d}{2})\;\left((k_1\cdot p)k^{2}_1-(k^{2}_1)^2\right)\;\Delta^{\frac{d}{2}-2}\right],
\end{eqnarray}

\noindent
where $\Delta = m^2 - x(1-x)k_{1}^{2}$. Continuing our calculation, we
would introduce two new parametric integrations as there are now three
different denominators (we take $1/\Delta$ as a new denominator)
depending on $k_1$ in each of the terms of the above expressions.
However, as the result does not depend on $m$ and we are looking only
for the pole part of the amplitudes, we can speed up the calculation
by modifying the dependence on $m$ of some denominators. For example,
without changing the final result, we can replace the first term on
the right hand side of (\ref{I6}) by
\begin{equation}\label{I8}
 {\cal T}\frac{\pi^{(2-\frac{d}{2})}}{2^{d-3}}\int_{0}^{1}dx\int\frac{d^{d}k_{1}}{(2\pi)^{d}}\;\frac{\Gamma(1-\frac{d}{2})\;\Delta^{\frac{d}{2}-1}}{(k^{2}_{1}-m^{2})^2}
\end{equation}
Similarly, in the computation of $B_3(a)$ one can set $m=0$ in the
expression for $\Delta$ so that one has to use only one parametric
integral. Following this recipe, after integrating in $k_1$ we obtain
($a=1-y-yx (1-x)$),
\begin{eqnarray}\label{I9}
\frac{A_3(a)}{\epsilon}&=& {\cal T}\frac{ \pi^{2-d}}{2^{2 d - 3}}\Gamma[3-d]
\int_{0}^{1}dx \int_{0}^{1}dy \, (1-y) 
\nonumber \\
&\times& [\frac{(1-2 y) m^2}{a}]^{d-3}
\left [ \frac{y^{-d/2}}{a^{3-d/2}} +\frac{d}{2} ( 1 +x - x^2) 
 \frac{y^{1-d/2}}{a^{4-d/2}}\right ]\nonumber\\
&=& -\frac{5i}{8 \epsilon}
\end{eqnarray}
\noindent
and
\begin{eqnarray}\label{I10}
\frac{B_3(a)}{\epsilon}&=&-{\cal T}\frac{ \pi^{2-d}}{2^{2 d - 3}}\Gamma[3-d]
\int_{0}^{1}dx \int_{0}^{1}dy \, \frac{(1-y)^{2-d/2}(y m^2-y(1-y)p^2)^{d-3}}
{[x(x-1)]^{1-d/2}}  \nonumber \\ 
&\times & \left [ \frac{(1-5 y)\Gamma[1-d/2]}{\Gamma[3-d/2]}- 2 
\frac{\Gamma[2-d/2]\Gamma[5-d/2]}{\Gamma[4-d/2]^2} (1-y) (1- 7 y)\right]
\nonumber \\
&=& - \frac{i}{24 \epsilon}.
\end{eqnarray}

Now, for the remaining graphs of order $\alpha^2$, we have,

Figure \ref{fig2}$(b):$
\begin{eqnarray}\label{I11}
I_3(b)&=&-4i\pi^2 
\;\epsilon_{\mu\nu\lambda}\;\epsilon_{\alpha\beta\rho}{\cal T}\int
\frac{d^dk_1}{(2\pi)^d}\frac{d^dk_2}{(2\pi)^d}\;k^{\lambda}_{1}\;k^{\rho}_{2}
\nonumber \\
&\times& \frac{\left[\gamma^{\mu}(\not \! p-\not \!
k_{1}+m)\gamma^{\alpha}(\not \! p-\not \! k_{1}-\not \!
k_{2}+m)\gamma^{\beta}(\not \! p-\not \! k_1+m)\gamma^{\nu}\right]}{[(p-k_{1})^2-m^2]^2\;[(p-k_1-k_2)^2-m^2]\;k^{2}_{1}\;k^{2}_2}
\nonumber \\
&=&\left(m\;A_3(b)+\not \! p\;B_3(b)\right)\frac{1}{\epsilon}
\end{eqnarray}
 Following the same steps as in previous case, we obtain
\begin{eqnarray}\label{I12}
\frac{A_3(b)}{\epsilon}&=&\frac{1}{2m} {\rm Tr}\;I_3(b)|_{p=0}=
- \frac{i}{4\epsilon}  \\
\frac{B_3(b)}{\epsilon} &=&\frac{1}{2p^2}\;Tr(I_3(b)\not \! p)=
- \frac{i}{12\epsilon}
\end{eqnarray}

Figure \ref{fig2}$(c):$
\begin{eqnarray}\label{I13}
I_3(c)&=&-4i\pi^2 
\;\epsilon_{\mu\beta\lambda}\;\epsilon_{\alpha\nu\rho}{\cal T}\int
\frac{d^dk_1}{(2\pi)^d}\frac{d^dk_2}{(2\pi)^d}\;k^{\lambda}_{1}\;k^{\rho}_{2}
\nonumber \\
&\times& \frac{\left[\gamma^{\mu}(\not \! p-\not \!
k_{1}+m)\gamma^{\alpha}(\not \! p-\not \! k_{1}-\not \!
k_{2}+m)\gamma^{\beta}(\not \! p-\not \! k_2+m)\gamma^{\nu}\right]}
{[(p-k_{1})^2-m^2]\;[(p-k_1-k_2)^2-m^2]\;[(p-k_{2})^2-m^2]\;k^{2}_{1}\;k^{2}_2}
\nonumber \\
&=&\left(m\;A_3(c)+\not \! p\;B_3(c)\right)\frac{1}{\epsilon}.
\end{eqnarray}
The computation of this expression is a bit more complicated because one
has to introduce three Feynman parametric integrals. The final result is,
nevertheless, simple,
\begin{eqnarray}\label{I14}
A_3(c) &=& -\frac{3i}{8}\\
B_3(c) &=&  \frac{i}{24}
\end{eqnarray}

Collecting these results, we obtain
\begin{eqnarray}
A_3 &=& A_3(a)+A_3(b)+A_3(c)= -\frac{5 i}{4}\nonumber\\
B_3 &=& B_3(a)+B_3(b)+B_3(c)= -\frac{ i}{12}\label{I15}
\end{eqnarray}
2. Order $ g \alpha$ graphs. There are six diagrams which have been drawn
in Fig. \ref{fig3}. They give

Figures \ref{fig3}$(a)$ and \ref{fig3}$(b)$. Both diagrams have the
structure of a product of two one loop graphs. The corresponding
dimensionally regularized amplitudes does not have a pole at $d=3$.

Figure \ref{fig3}$(c)$. Actually, this diagram does not contribute because the
corresponding analytic expression is finite. Indeed, we have
\begin{eqnarray}\label{I16}
I_4(c) &=& {\cal T}\frac{\partial \phantom {a}}{\partial m}\int 
\frac{d^dk_1}{(2\pi)^d}\frac{d^dk_2}{(2\pi)^d}\; \frac{{\rm Tr} [\gamma^\mu (\not \!
k_2 + m)\gamma^\nu(\not \! k_2+\not \! k_1+ m)] \epsilon_{\mu\nu\rho}}
{(k_{2}^{2}-m^2) ((k_1+k_2)^2-m^2)} \frac
{k_{1}^{\rho}}{k_{1}^{2}} \nonumber\\
&=& - 4 i{\cal T}\frac{\partial \phantom {a}}{\partial m}\int\frac{d^dk_1}{(2\pi)^d}\frac{d^dk_2}{(2\pi)^d}\;\frac{m}{(k_{2}^{2}-m^2) ((k_1+k_2)^2-m^2)}=0
\end{eqnarray}
since the integral in the second equality has the structure of a product of
two one loop integrals.

Figure \ref{fig3}$(d)$. The same reasoning can be applied to this
situation since no external momentum flows through the diagram. We
conclude that there is not a pole term.

Figure \ref{fig3}$(e)$. By Furry's theorem this diagram cancels with its charge
conjugated partner.

Figure \ref{fig3}$(f)$. We have the following contribution
\begin{eqnarray}\label{I17}
I_4(f)&=& 4\pi \epsilon_{\mu\nu\lambda}\;{\cal T} \int
\frac{d^{d}k_1}{(2\pi)^d}\frac{d^{d}k_2}{(2\pi)^d}\;\gamma^{\mu}(\not
\! p-\not \! k_1+m)k^{\lambda}_{1}
\nonumber \\
&\times & \frac{(\not \! k_2-\not \!
k_1+m)\gamma^{\nu}(\not \! k_2+m)}{[(p-k_1)^2-m^2]\;(k^{2}_2-m^2)\;((k_2-k_1)^2-m^2)\;k^{2}_1}\nonumber\\
&=& -i\frac{(m^2 A_4(f)+ m\not \! p B_4(f))}{\epsilon}
\end{eqnarray}
from which one obtains
\begin{eqnarray}
A_4&=& A_4(f) =  \frac{9}{16 \pi}\nonumber\\
B_4&=&B_4(f) = \frac{1}{8\pi}\label{I18}
\end{eqnarray}

3. Order $g^2$ graphs. There are only the two diagrams shown in Fig $f_5$. We
get,
\begin{eqnarray}\label{I19}
I_5 &=&-4 i{\cal T}\int\frac{d^{d}k_2}{(2\pi)^{d}}\frac{d^{d}k_1}{(2\pi)^{d}}
\frac{\not\! p + \not\! k_1+ m}{[(p+k_1)^{2}-m^{2}](k_{2}^{2}-m^{2})[(k_2-k_1)^{2}-m^{2}]}\nonumber\\
&\times& \{(\not \! k_2+ m)(\not \!k_2-\not\! k_1+ m)- {\rm Tr}[(\not \! k_2+ m)(\not \!k_2-\not\! k_1+ m)]\}\nonumber \\
&=& -\frac{(m^3 A_5+ m^2 \not \! p B_5)}{\epsilon},
\end{eqnarray}

\noindent
where the two terms on the second equality refers to the graphs
\ref{fig4}$(a)$ and \ref{fig4}$(b)$, respectively. After a lengthy
calculation one determines,
\begin{eqnarray}
A_5 &=&- \frac {3i}{16 \pi^2}\nonumber\\
B_5 &=&- \frac{5i}{48 \pi^2}\label{I20}
\end{eqnarray}
\section{}\label{appendixB}
In this appendix we will discuss the calculation of the pole part of
the four point vertex function which is needed for fixing the
renormalization group beta function. Actually, since all we need is the
constant part of the residue, the calculation of the relevant graphs
will be done at zero external momenta. 

The first observation is that, up to the order we are interested, i.
e., third order, there are too many graphs. To be systematic, we will
separate them accordingly they have or have not closed fermionic
loops. Moreover, if they do not posses fermionic loops we group
them accordingly the number of CS or auxiliary GN lines linking the two fermion
lines crossing the diagram. Many diagrams cancel because of Furry's
theorem; this is the case if there is a fermionic loop with an odd
number of attached CS lines. Other diagrams have the structure of a
product of two one loop graphs and therefore are finite.  We shall not
consider these two types of graphs any longer. The anti-symmetrized
amplitude for a graph $\gamma$ has the generic structure of a product,
$(A\otimes B)\,C$, where A and B refers to the propagators and vertices
associated to the two fermion lines and C to the others factors
($\otimes$ indicates the anti-symmetrized direct product).  Using this
notation, one can verify that
\begin{equation}\label{B1}
\mbox{ Pole Part of} \int d^dk_1\, d^dk_2\,\, (A\otimes B) C= \frac{\cal T}{2}\int d^dk_1\, 
d^dk_2(
{\rm Tr}[A]\,\,{\rm Tr}[B]-{\rm Tr}[AB])C,
\end{equation}
 
For example, from the analytic expression for the graph shown in Fig.
\ref{fig5},
\begin{eqnarray}
\lambda\int  d^dk_1\, d^dk_2 [\gamma^\mu S_F(k_2)]\otimes [\gamma^\nu S_F(-k_2)]
{\rm Tr}[S_F(k_1) S_F(k_1-k_2)] \epsilon_{\mu\nu\lambda} 
\frac{k_{2}^{\lambda}}{k^{2}_{2}} ,
 \end{eqnarray} \label{B2}
where $\lambda$ is a combinatorial factor, we determine $A$, $B$ and $C$ as
\begin{eqnarray}\label{B3}
A&=& [\gamma^\mu S_F(k_2)]\\
B&=& [\gamma^\nu S_F(-k_2)]\\
C&=& \lambda\epsilon_{\mu\nu\lambda}\frac{k_{2}^{\lambda}}{k^{2}_{2}}
\end{eqnarray}

The results for the pole parts are summarized in the Tables $A$ and
$B$, which corresponds to the two cases mentioned above. In the table
A are listed the results from graphs with one closed fermionic loop;
these have been arranged accordingly the number of CS vertices in the
loop. Table \ref{appendixB} exhibits the pole part of graphs without fermionic loop.
They have been collected into types (i,j), where i and j are the number 
auxiliary GN and CS lines,respectively,  linking the two fermion
lines crossing the diagram. Notice that there are not contribution from
graphs of type (0,1) since they are not proper.
Figure \ref{fig6}  furnishes examples of each one
of these sets of diagrams. The final result for each order is obtained
by summing the corresponding entries in each table. Thus we have,
\begin{equation}\label{B4}
C_3=-\frac{7}{2} \qquad \qquad C_4=\frac{11i}{2\pi}\qquad \qquad C_5=\frac{13}{2\pi^2}
\end{equation}

\begin{center}
ACKNOWLEDGMENTS  
\end{center}

This work was supported in part by Conselho Nacional de
Desenvolvimento Cient\'\i fico e Tecnol\'ogico (CNPq) e Funda\c c\~ao de
Amparo \`a Pesquisa do Estado de S\~ao Paulo (FAPESP).

\begin{figure}
\caption{Feynman rules for the interaction vertices. Continuous and wavy lines
represent the fermion and vector propagators, respectively.} 
\label{fig1}
\end{figure}
\begin{figure}
\caption{Order $\alpha^2$ graphs contributing to the two point function.} 
\label{fig2}
\end{figure}
\begin{figure}
\caption{Fermionic self-energy graphs of order $g\alpha$.} 
\label{fig3}
\end{figure}
\begin{figure}
\caption{Order $g^2$ fermionic self-energy graphs.} 
\label{fig4}
\end{figure}
\begin{figure}
\caption{Example of a two-loop diagram.} 
\label{fig5}
\end{figure}
\begin{figure}
\caption{Diagrams illustrating the various classes of graphs in the four
point vertex function} 
\label{fig6}
\end{figure}

\begin{table}
\caption{Pole part for four legs graphs with a closed fermionic loop.\label{tb1}}
\begin{tabular}{lll|}
{Order of perturbation}&{Number of diagrams}&{Pole part}\\
\tableline
$g\alpha^2$&12&${-i}/{2\epsilon}$\\
$g^2\alpha$&7&$-4i/\pi\epsilon$\\
$g^3$&8&$ 4i/\pi^2\epsilon$\\
 \end{tabular}
\end{table}

\begin{table}
\caption{Pole part for four legs graphs without  closed fermionic loops.\label{tb2} The first column lists different types of diagrams (i,j), where i and j
are the number of GN and CS lines joining the two fermion lines crossing
the graph; the digit in parenthesis after the pole parts is the number
of contributing graphs.}
\begin{tabular}{llll|}
{Diagram type}&{Order $g \alpha^2$}&{Order $g^2 \alpha$}&Order $g^3$\\
\tableline
(0,2)& ${i}/{2\epsilon}$\phantom{a} (8)&---&---\\
(1,0)&$-5i/2\epsilon$\phantom{a} (12)& $3i/\pi \epsilon$ (18) & $-3 i/4\pi^2 \epsilon$ (6)\\
(1,1)&$8 i/\epsilon$ (24)&$4i/\pi\epsilon$ (8)&---\\
(1,2)&$-2 i/\epsilon$ \phantom{a}(12)&---&---\\
(2,0)&---&$-i/\pi\epsilon$ (14)& $i/\pi^2\epsilon$ (12)\\
(2,1)&---&$-15 i/2 \pi \epsilon$ (9) &---\\
(3,0)&---&---&$5i/4\pi^2\epsilon$ (4)\\
\end{tabular}
\end{table}


\begin{thebibliography}{99}
\bibitem {o1} We use natural units $(c=\hbar=1)$ and our metric is
$g_{00}=-g_{11}=-g_{22}=1$. The fully antisymmetric tensor
$\epsilon^{\mu \nu \lambda}$ is normalized such that
$\epsilon^{012}=1$ and we define $\epsilon^{ij} \equiv
\epsilon^{0ij}$.  Repeated greek indices sum from 0 to 2, while
repeated Latin indices from the middle of the alphabet sum from 1 to
2.  For the $\gamma$-matrices we adopt the representation
$\gamma^{0}=\sigma^{3},\gamma^{1}=i\sigma^{1},\gamma^{2}=i\sigma^{2}$,
where $\sigma^{i}, i=1,2,3,$ are the Pauli spin matrices.
\end{thebibliography}
\end{document}